\documentclass[twocolumn,english,reprint]{revtex4}
\usepackage[T1]{fontenc}
\usepackage[latin9]{inputenc}
\usepackage{textcomp}
\usepackage{graphicx}

\makeatletter
\@ifundefined{textcolor}{}
{%
 \definecolor{BLACK}{gray}{0}
 \definecolor{WHITE}{gray}{1}
 \definecolor{RED}{rgb}{1,0,0}
 \definecolor{GREEN}{rgb}{0,1,0}
 \definecolor{BLUE}{rgb}{0,0,1}
 \definecolor{CYAN}{cmyk}{1,0,0,0}
 \definecolor{MAGENTA}{cmyk}{0,1,0,0}
 \definecolor{YELLOW}{cmyk}{0,0,1,0}
}

\makeatother

\usepackage{babel}
\begin{document}

\title{Magnetic domain-wall motion study under an electric field in a Finemet\textregistered{}
thin film on flexible substrate}

\author{Ngo Thi Lan$^{1}$, Silvana Mercone$^{1}$%
\footnote{Email: silvana.mercone@univ-paris13.fr%
}, Johan Moulin$^{2}$, Anouar El Bahoui$^{1}$, Damien Faurie$^{1}$,
Fatih Zighem$^{1}$, Mohamed Belmeguenai$^{1}$ and  Halim
Haddadi $^{1}$}

\affiliation{$^{1}$ Laboratoire des Sciences des Procédés et des Matériaux, CNRS-Université
Paris XIII, 93430 Villetaneuse, France}

\affiliation{$^{2}$ Institut d'Electronique Fondamentale, UMR 8622 Université
Paris Sud / CNRS, Orsay, France}
\begin{abstract}
Influence of applied in-plane elastic strains on the static magnetic
configuration of a 530 nm magnetostrictive FeCuNbSiB (Finemet\textregistered{})
thin film. The in-plane strains are induced via the application of
a voltage to a piezoelectric actuator on which the film/substrate
system was glued. A quantitative characterization of the voltage dependence
of the induced-strain at the surface of the film was performed using
a digital image correlation technique. MFM images at remanence ($H=0$
Oe and $U=0$ V) clearly reveal the presence of weak stripe domains.
The effect of the voltage-induced strain shows the existence of a
threshold value above, which the break of the stripe configuration
set in. For a maximum strain of $\varepsilon_{XX}\sim0.5\times10^{-3}$
we succeed in destabilizing the stripes configuration helping the
setting up of a complete homogeneous magnetic pattern.
\end{abstract}


\maketitle

\section{Introduction }

During the last decades technological progress has been driven predominantly
by the modern information and communication technology. The steadily
increasing data output and functionality of devices has required an
ongoing miniaturization of their structural elements. In the design
and manufacturing of actual microelectronic and microelectromechanical
systems (MEMS), thin metallic films play an important role \cite{Niarchos2003,Craighead2000}.
Thin films can show continuous geometry for which lateral dimensions
are much higher than thickness (typically less than one micron), or
show complex geometry (wires or dots arrays) for which one or two
lateral dimensions and thickness are of same order. Usually, thin
films are deposited on substrates, which are several ten times thicker.
In the case of flexible substrates (generally polymers) used in stretchable
electronics, the thin films are usually submitted to mechanical stresses
due to the curvature of the whole system. Obviously, these stresses
may have an important effect on the magnetic properties, especially
on the effective magnetic anisotropy \cite{Dai2012,Shin2012}. Magnetic
devices fabricated on compliant substrates such as polymers are believed
to have great potential for applications due to their mechanical fl{}exibility,
enhanced durability and lightweight compared with those on a rigid
substrate. Moreover, it has been found that the specifi{}c nature
of polymer substrates has non-negligible consequence on the magnetic
properties of deposited fi{}lms, which leads to a different performance
from a fi{}lm deposited on rigid substrate. Due to their stress sensitivity,
the magnetic properties of magnetostrictive fi{}lms on fl{}exible
substrate can be tailored by the stress \cite{Xu2010,Ludwig2002,Ozkaya2008,Iakubov2012,Uhrmann2006}.
In this context, a few dedicated techniques have been developed to
study the magnetelastic behavior of thin films. Concerning magnetic
films deposited on flexible substrates, recent papers have reported\textit{
in situ} characterizations by FerroMagnetic Resonance (FMR) \cite{Zighem2013},
Magneto-Optical Kerr Effect\cite{Ozkaya2008} and Giant MagnetoResistance
(GMR) measurements \cite{Uhrmann2006,Barraud2010,Melzer2011}. The
purpose of these studies is especially to demonstrate the feasibility
of a magnetostrictive sensor on a polymer substrate. The main advantage
of polymer materials is obviously their high flexibility, which cannot
be achieved by crystalline materials, and their relative low cost.

In the present paper, the influence of an applied elastic strain on
the magnetic domain of a Finemet\textregistered{} film/Kapton\textregistered{}
substrate is investigated. The external loading is applied thanks
to a piezoactuator \cite{Pi_actuator} on which our system is glued,
as already shown in previous studies \cite{Zighem2013,Brandlmaier2009,Brandlmaier 2008}.
A quantitative study of the voltage-induced elastic in-plane strains
is performed thanks to a Digital Image Correlation (DIC) technique
while the evolution of magnetic domain structures is probed by Magnetic
Force Microscopy (MFM).

\begin{figure}[h]
\centering{}\includegraphics[bb=30bp 290bp 740bp 580bp,clip,width=8cm]{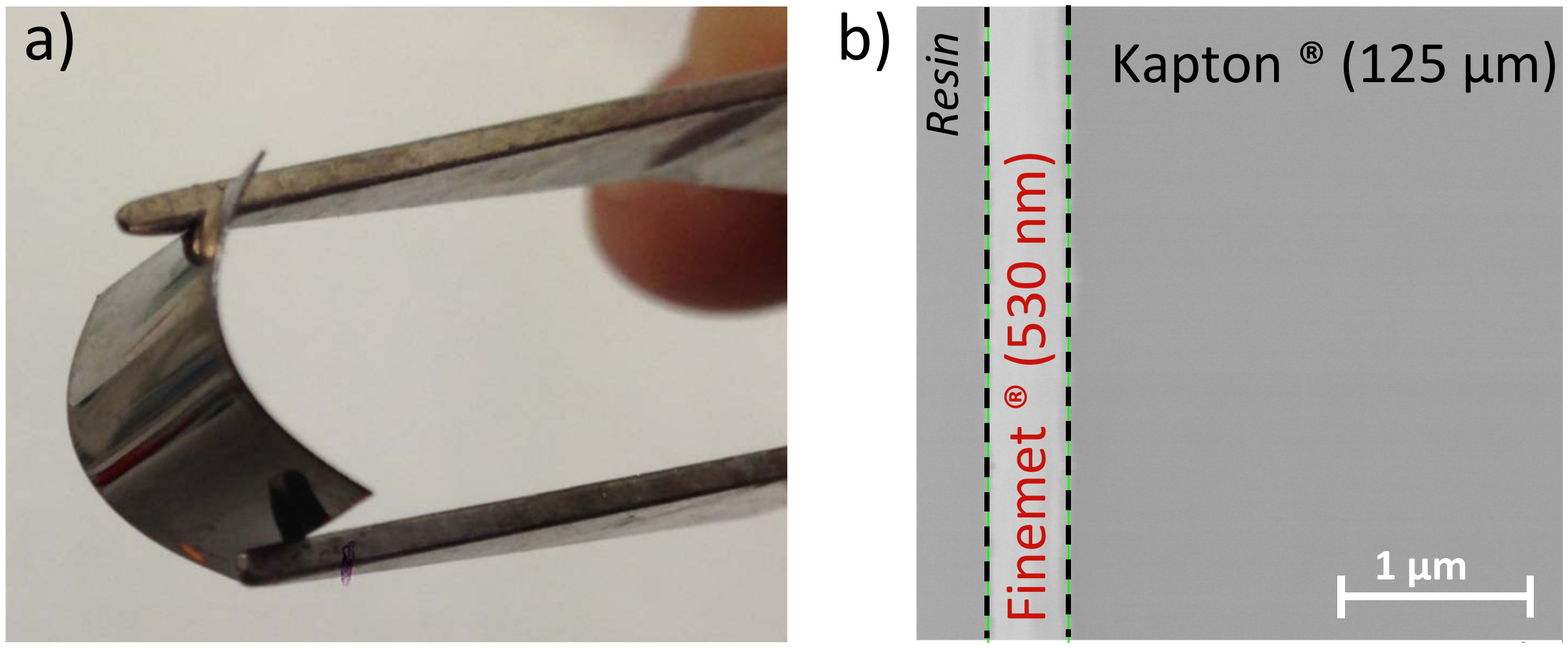}
\caption{a) Image of the flexible Finemet/Kapton\textregistered{} system. b)
Cross-section of the Finemet/Kapton\textregistered{} system obtained
by Scanning Electron Microscope (SEM). A Finemet film thickness of
about 530 nm is estimated.}
\end{figure}

\section{Experimental}

The amorphous Finemet\textregistered{} film has been deposited on
Kapton\textregistered{} substrate (125 \textmu{}m) (shown on figure
1-a) by RF sputtering with the following conditions : residual pressure
in the range of 10-7 mbar, working Ar pressure 40 mbar and RF power
250 W. A thin film of titanium was deposited under the layer in order
to increase adhesion to the Kapton and over the layer in order to
protect from oxidation. The composition of the target was Fe$_{73.5}$Cu$_{1}$Nb$_{3}$Si$_{15.5}$B$_{7}$.
The film composition has been characterized by EDS and is close to
the one of the target. The film thickness (530 nm) has been estimated
by Scanning Electron Microscopy equipped with a Field Electron Gun
(SEM-FEG) (see the cross-section view on figure 1-b).

\begin{figure*}
\begin{centering}
\includegraphics[bb=30bp 340bp 720bp 575bp,clip,width=12cm]{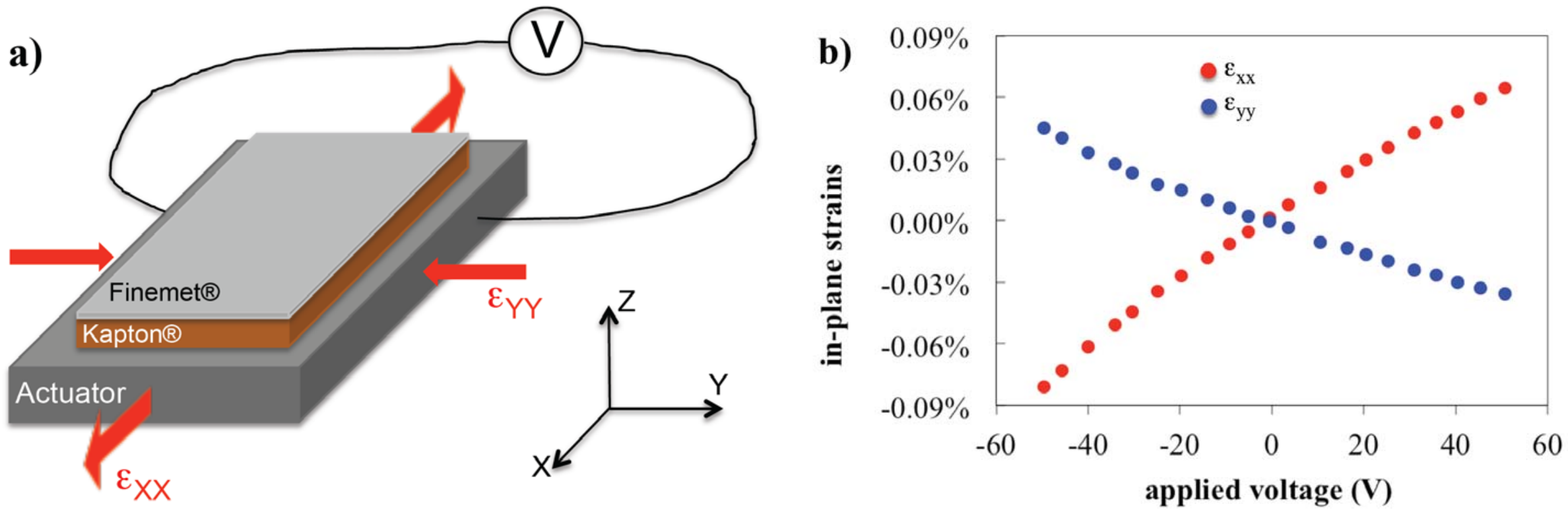}
\par\end{centering}

\caption{a) Sketch of the artificial FeCuNbSiB/piezoelectric heterostructure.
b) Quantitative characterization of the voltage dependence of the
induced-strain at the surface of the film was performed using a digital
correlation image technique. }
\end{figure*}

After deposition, the flexible film/substrate system was glued onto
a piezoelectric actuator. Zighem et al. \cite{Zighem2013} have already
shown that a compliant substrate allows to have a nearly $100\%$
strain transmission in between the piezoelectric actuator and the
film. In the present work, a quantitative characterization of the
voltage dependence of the induced-strain at the surface of the film
was performed using Digital Image Correlation technique \cite{Haddadi2012}.
Magnetic Force microscopy (MFM) measurements are performed by using
a standard Veeco D3100 microscope. In order to provide reasonables
sensitivity and resolution, standard hard magnetic tip (PPP-MFMR tips)
are used. The hard magnetic coating of this tip has approximately
300 Oe coercivity and a 300 emu.cm$^{-3}$ remanence magnetization
which allow high magnetic contrast. This latter characteristic is
mandatory for weak magnetic stray field detection and high lateral
resolution domain imaging (i.e. around 50 nm). The MFM has been used
as an \textit{in situ} probe of the static magnetic domains configuration
under the influence of applied in-plane elastic strains. Before this,
in order to fully characterize the ferromagnetic film, the ground
state ($U=0$ V) of the system is characterized. We performed standard
contact Atomic Force Microscopy (AFM) for a roughness check of the
surface of the ferromagnetic film and static magnetization measurements
at room temperature for static magnetic behavior.

\section{Results and discussion}

The sketch of the studied artificial multiferroic system made by the
530 nm Finemet\textregistered{} ferromagnetic film deposited onto
the compliant polyimide substrate (Kapton\textregistered{}) glued
on a piezoelectric device is shown in figure 2-a. This latter allows,
via the application of a voltage, to induce in-plane strains ($\varepsilon_{XX}$
and $\varepsilon_{YY}$) to the film/substrate, thanks to the choice
of the Kapton\textregistered{} substrate which avoids clamping effects
leading to low transmission of strains from the actuator to the film/substrate
system \cite{Brandlmaier 2008}. The DIC measurements have been performed
by varying the external voltage from +50 V to -50 V. The measured
in-plane strains ($\varepsilon_{XX}$ and $\varepsilon_{YY}$ ) in
this range of voltages are presented in Figure 2-b. The reference
image for the determination of the in-plane strains has been taken
at $U=0$ V (after applying a voltage of $+50$ V in order to avoid
hysteresis effects due to the piezoelectric material). In these conditions,
a maximum (resp. minimum) value of 0.065\% (resp. -0.035\%) for $\varepsilon_{XX}$
(resp. for $\varepsilon_{YY}$) is found at $U=+50$ V. One can note
that the values obtained at $U=-50$ V are slightly higher (-0.085\%
for $\varepsilon_{XX}$ and 0.045 for $\varepsilon_{YY}$) than the
ones obtained at $U=+50$ V. This behavior is due to the non-linear
and hysteretic behavior of the piezoelectric actuator. However, we
have verified that no variations of these values are observed after
several $U=+50$ V to $U=-50$ V sweeps (note that the sweep $U=-50$
V to $U=+50$ V is not shown here). In addition, one can note that
the sketch presented in figure 2-a corresponds to a positive applied
voltage where the magnetic film is tensily stressed (when a negative
voltage is applied, the film is compressively stressed).

The influence of voltage-induced in-plane elastic strains on the magnetic
domain of the film has been probed by MFM. The presence of magnetic
domains has been clearly identified at zero applied voltage by MFM
(see figure 3-a). This kind of micromagnetic configuration appears
in magnetic film presenting an out-of plane magnetic anisotropy (uniaxial
in the present case) under specific conditions. Indeed, the magnetic
domains presented in figure 3-a appear only if the thickness of the
film is higher than a critical thickness $t_{C}=2\pi\sqrt{\frac{A}{K_{out}}}$
where $A$ is the exchange stiffness and $K_{out}$ is the out-of-plane
magnetic anisotropy constant and if $K_{out}$ is smaller than the
demagnetizing energy contribution (here $\sim2\pi M_{S}^{2}$) \cite{Talbi2010}.
The thickness of the studied film is around 530 nm which is higher
than the calculated critical thickness.

The MFM tip was magnetized along its axis using an external magnet
with field $>1$ kOe in order to have a magnetic moment directed along
the apex-base axis of the pyramidal nanometric tip. This moment is
a sensitive \textit{in situ} probe of the stray field coming from
the surface of the studied film and always perpendicular to it. As
a simple approximation, one can consider a dipole-dipole interaction
between the tip and the stray field coming from changes in the micromagnetic
configuration of the film. Operating in tapping mode in air, it is
easy to find a direct connection between the shift in phase of the
eigen resonance frequency of the tip and the force gradient of the
dipole-dipole interaction:

\[
\Delta\phi=\frac{-Q}{k}\frac{\partial F}{\partial z}
\]

In this latter expression $k$ is the cantilever spring constant and
$Q$ is the vibrating system quality factor. In these conditions,
attractive (resp. repulsive) forces with a positive (resp. negative)
gradient lead to a negative (positive) phase-lag: dark contrast (resp.
white contrast).

\begin{figure*}
\begin{centering}
\includegraphics[bb=50bp 280bp 480bp 575bp,clip,width=12cm]{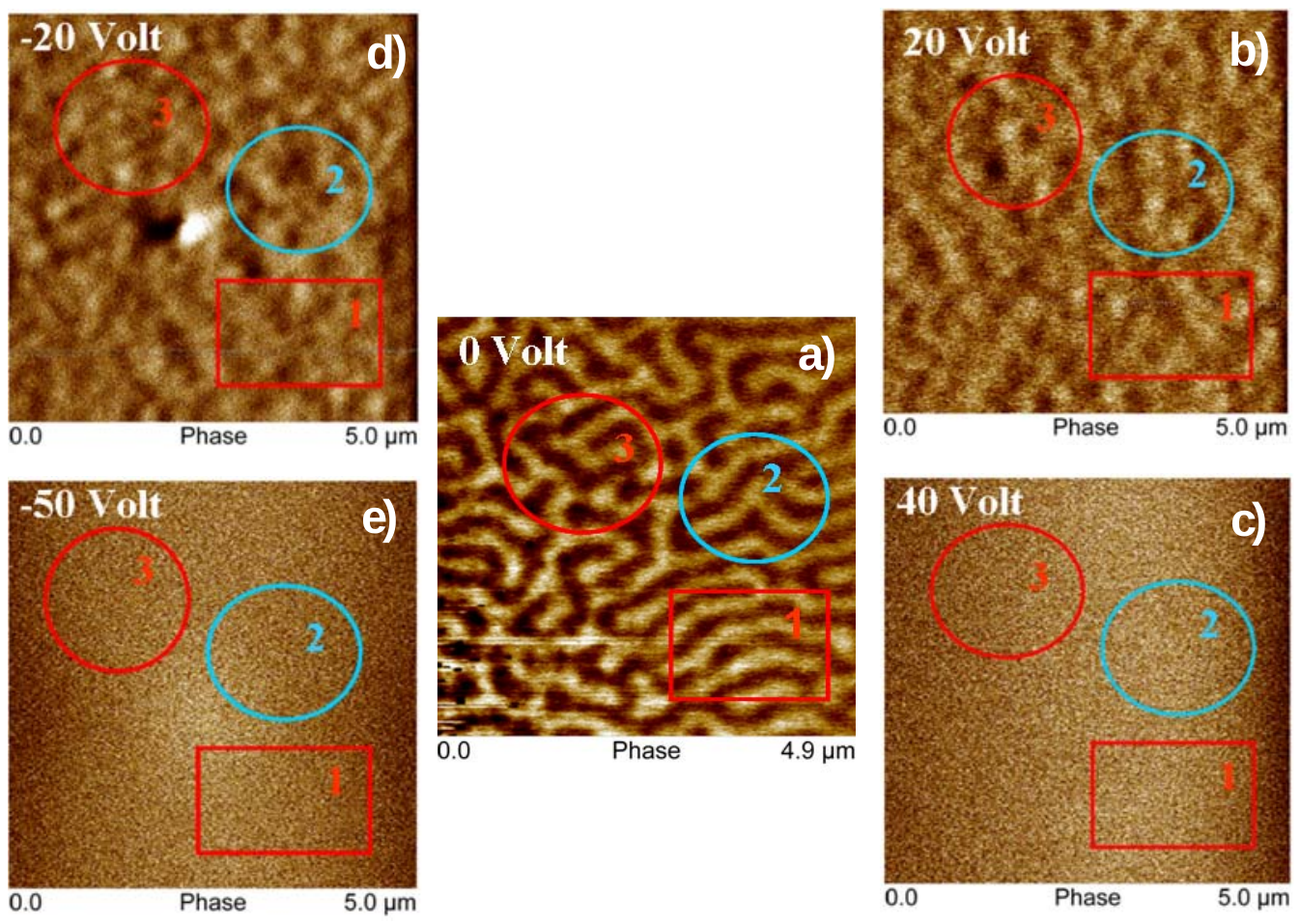}
\par\end{centering}

\caption{Strain-induced modification of magnetic domain patterns in the film
\textit{via} an applied voltage. MFM phase images at (a) $U=0$ V
; (b) $U=+20$ V ; (c) $U=+40$ V ; (d) $U=-20$ V ; (e) $U=-50$
V. }
\end{figure*}

\begin{figure*}
\begin{centering}
\includegraphics[bb=50bp 100bp 800bp 565bp,clip,width=12cm]{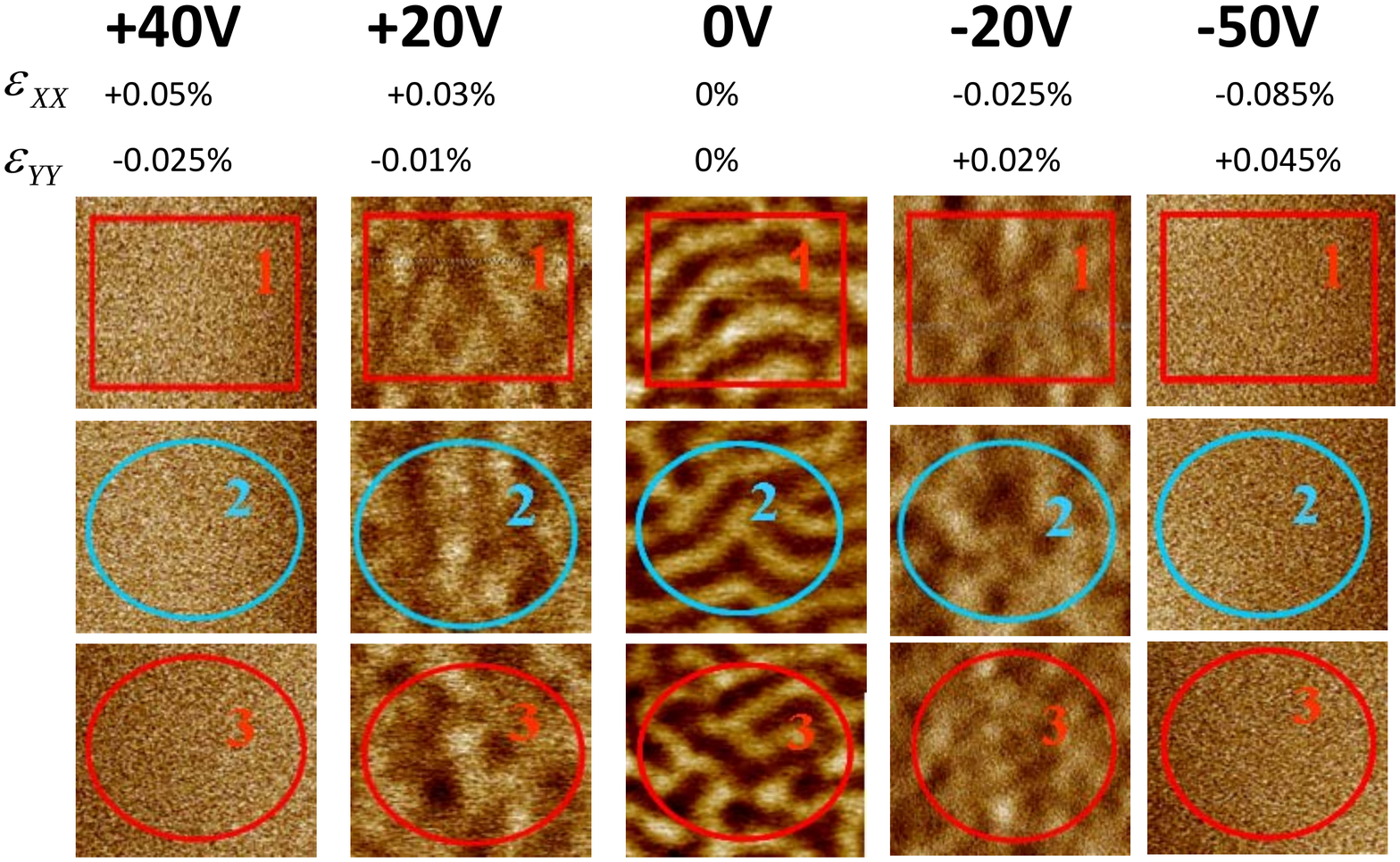}
\par\end{centering}

\caption{Zoom in of 3 different zones showing various changes of magnetic stripe-domain
patters : (zone 1) from curved domains to maze-like pattern ; (zone
2) from stripe domaine to maze-like pattern and (zone 3) melting of
the maze-like pattern.}
\end{figure*}

Figure 3 shows the images representation of this phase changes over
the observed film surface for different voltage-induced in-plane strains.
Tapping voltage set point, drive amplitude of the tip and lift scan
distance has been adjusted all along measurements in order to optimize
contrast in the phase image variations and to prevent unwanted local
magnetization effect of the magnetic film by the MFM tip. Also lift
scan distance has been optimized in order to avoid tip demagnetization
effect by the stray field of the film. Typical set point values were
between 0 and 1.5 Volt and lift distances in the range of 80 to 150
nm. Typical scan area was of $5\times5$ \textmu{}m$^{2}$ at a scan
rate between 0.3 and 0.5 Hz. Both topography and magnetic images were
obtained for each strained configuration. In first approximation,
black and white regions represent opposite magnetization direction
giving place to opposite sign of the dipole-dipole interaction between
the tip and the stray field coming from the surface sample. The use
and fabrication of the studied heterostructure allows the application
of relatively high in-plane strains which lead to drastic change of
the micromagnetic configuration inside the film: from ``stripe''-domains
(see Figure 3-a) to a maze-like pattern (see Figures 3-b and 3-d)
and finally to a ``macrospin'' configuration where no magnetic contrast
is observed (see Figures 3-c and 3-e). Specifically, Figure 3b-d show
that the domain structure is transformed into a maze-like complex
pattern all over the studied surface when a voltage of $+20$ V and
$-20$ V is applied.

These changes are due to the voltage-induced strain control of the
magnetoelastic anisotropy of the Finemet\textregistered{} film. In
first approximation, this magnetoelastic anisotropy contribution can
be view as a static magnetic field applied in the plane of the film.
It should noted that very small induced strains are sufficient to
modify the micromagnetic configurations into an apparent uniform configuration.
This extreme sensivity of the magnetic configuration is due to the
non-negligible magnetostriction of the studied film. In order to strengthen
these observations, we show here in Figures 4 a zoom-in of specific
boxed regions in figures 3 (box 1, 2 and 3). It is clear from all
the three zones chosen (i.e. bended+curved domains=zone 1, stripe
parallel domains=zone 2 and puzzled-like domains=zone 3) that after
the application of +20 V strains big displacement of domain-walls
happened letting a different maze-like pattern to set in. At -20 V,
this latter presents the same main features but with opposite magnetic
domain orientations (i.e. white and black domains are exchanged).
After the application of +40 V exactly as for -50 V (see Figures 3
and 4) the domain pattern is fully supressed and an apparent uniform
magnetic configuration set in. In this latter phase images, the slight
change in colors have to be ascribed to the sample roughness, which
can locally influence the magnetic stray field.

\section{Conclusions}

In conclusion we investigated the influence of an applied elastic
strain on the magnetic domain of a Finemet\textregistered{} film/Kapton\textregistered{}
substrate. The applied elastic strains were controlled by applying
voltage to a piezo-actuator on which the Finemet\textregistered{}/Kapton\textregistered{}
where glued. The amount of strains was measured by Digital Image Correlation
while using Magnetic Force Microscopy (MFM) we probed changes in magnetic
domain structures. MFM images at remanence (H=0Oe and U=0V) clearly
reveal the presence of weak stripe domains. We have found that magnetic
domains in Finemet\textregistered{} alloy films showed high mobility
upon small applied strains. A threshold value of the electric field
applied has been observed. Above this latter the break of the stripe
configuration sets in by bending, curving and branching of domains
and gives way to a maze-like configuration. For a maximum strain of
$\varepsilon_{XX}\sim0.5\times10^{-3}$ the stripes configuration
is destabilized and a complete homogeneous magnetic pattern appeared.

\section{Acknowledgments}

The authors gratefully acknowledge the CNRS for their fi{}nancial
support in the framework of the \textquotedblleft{}PEPS INSIS\textquotedblright{}
program (``Ferroflex'' project). Ngo Thi Lan is grateful to the
Labex SEAM for her financial support during her stay in Paris 13th
University. This work was also partially supported by the French Research
Agency (project ANR 2010 JCJC 090601 entitled \textquotedblleft{}SpinStress\textquotedblright{}).

\end{document}